\documentstyle[preprint,aps]{revtex}
\begin{document}
\draft
\preprint{\begin{tabular}{l}
\hbox to\hsize{March, 1995 \hfill SNUTP 95-024}\\[-3mm]
\hbox to\hsize{hep-ph/9503436 \hfill Brown-HET-990-Rev}\\[5mm] \end{tabular} }

\bigskip

\title{ A simple modification of the maximal mixing \\ scenario for three
light neutrinos }
\author{Kyungsik Kang}
\address{Department of Physics, Brown University\\ Providence,
Rhode Island 02912, USA}
\author{Jihn E. Kim}
\address{Center for Theoretical Physics and Department of
Physics\\ Seoul National University, Seoul 151-742, Korea}
\author{Pyungwon Ko }
\address{Department of Physics, Hong-Ik University \\ Seoul 121-791, Korea}
\maketitle
\begin{abstract}
We suggest a simple modification of the maximal mixing scenario 
(with $S_3$  permutation symmetry) for 
three light neutrinos. Our neutrino mass matrix has smaller 
permutation symmetry $S_{2}$ ($\nu_{\mu} \leftrightarrow \nu_{e}$), 
and is consistent 
with all neutrino experiments except the $^{37}$Cl experiment. 
The resulting mass eigenvalues for three neutrinos are 
$m_{1} \approx (2.55-1.27) \times 10^{-3}~~{\rm eV}, m_{2,3} \approx 
(0.71-1.43)~~{\rm eV}$ for $\Delta m_{LSND}^{2} = 0.5 - 2.0~{\rm eV}^2$.
Then these light neutrinos can account for
$\sim (2.4-4.8) \%~~(6.2-12.4 \%)$ of the dark matter for $h = 0.8~(0.5)$.   
Our model predicts the $\nu_{\mu} \rightarrow \nu_{\tau}$ oscillation 
probability in the range sensitive to the future experiments such as 
CHORUS and NOMAD.
\end{abstract}


\newpage
\narrowtext
 \tighten

The minimal standard model (MSM) has been highly 
successful in describing interactions 
among elementary particles from low energy up to $\sim 100$ GeV.  
The only possible exception may be various types of neutrino oscillation 
experiments.  There have been positive indications from large scale 
experiments for solar and atmospheric neutrinos that a certain  
amount of mixing between neutrino species may be present \cite{pdg}.   
The recent report from the LSND experiment at the laboratory scale provides 
us with another hint of such a possible neutrino mixing \cite{lsnd}.  
Since neutrinos in the MSM are exactly massless,
there can be no mixing among them, and  it is impossible to 
accommodate such neutrino mixing data in the framework of the MSM.  
This situation is rather encouraging, since it is at present the only place 
where we can grasp a hint of new physics beyond the MSM. 

In view of this, it is quite interesting to speculate  what type of 
neutrino mass matrix can fit all the  data from the  various types of  
neutrino  oscillation experiments.  It is our purpose to present one such  
mass matrix in this work.  Most analyses on the neutrino oscillation assume 
two neutrinos oscillating with one mass difference parameter, $\Delta m^2$. 
However, the
LSND experiment and the atmospheric and solar neutrino data
hint at least two mass difference parameters, requiring oscillations
among at least three neutrinos.  For oscillations with three
neutrinos, we have two mass differences, three real angles and one phase.
In order to simplify the analysis, a certain ansatz for the
mass matrix is required.  In this vein, we first briefly discuss 
the maximal mixing scenario for the neutrino sector. We then present 
our ansatz for the neutrino mass matrix as well as the numerical analyses 
to fit the atmospheric,  LSND, and
solar neutrino data from  GALLEX and SAGE.  In this work, we consider
oscillations among three neutrinos only, 
$\nu_{\alpha} \rightarrow \nu_{\beta}$. 

One of the popular ansatz for the neutrino mass matrix is the maximal mixing 
one (equivalent to a cyclic permutation symmetry among three generations) 
\cite{nussinov} :
\begin{eqnarray}
M_{maximal} = \left( \begin{array}{ccc}
             a & b & b^{*}   \\
             b^{*} & a & b \\
             b & b^{*} & a
           \end{array} \right),
\label{eq:maximalm}
\end{eqnarray}
with the mixing matrix $U$ given by 
\begin{eqnarray}
U_{maximal} = {1 \over \sqrt{3}}~\left( \begin{array}{ccc}
          \omega_{1} & \omega_{1} & \omega_{1}   \\
           \omega_{1} & \omega_{2} & \omega_{3}   \\
          \omega_{1} & \omega_{3} & \omega_{2}   
         \end{array} \right).
\label{eq:maximalu}
\end{eqnarray} 
Here, $\omega_{1,2,3}$ are three complex roots of $\omega^{3} = 1$ with
$\omega_1=1$.  
This ansatz was originally proposed in the neutrino sector \cite{nussinov},
and extended to the quark sector with a partial success in explaining 
the quark masses and the Kobayashi-Maskawa matrix elements \cite{scott}.

The maximal mixing scenario has many interesting features \cite{maximal}. 
For example,
the survival probability for a neutrino is independent of its flavor.
The $\nu_e$ survival probability has two plateaus, $5/9$ in the intermediate
step, and 1/3 for ${L\over E}\gg {1\over (\Delta m^2 _{ij})_{ \rm min}}$ 
through vacuum oscillations.
With vacuum oscillation, one cannot explain 
solar neutrino data from $^{37}$Cl and the Ga data simultaneously.              
Thus one must make a choice between the $^{37}$Cl and Ga experiments.
Here, we choose to interpret the Ga data (GALLEX and SAGE experiments)
through vacuum oscillation  and disregard the
$^{37}$Cl data \footnote{The matter oscillation effect, i.e., the MSW 
mechanism, has also been suggested to interprete this Homestake experiment
in Ref.~\cite{msw}. In this case, the relevant $\Delta m^2$ is around 
$\sim 10^{-4}~{\rm eV}^2$. Since we have only two $\Delta m^{2}$ around 
$\sim O(1)~{\rm eV}^2$ and $\sim 10^{-2}~{\rm eV}^2$, the MSW mechanism is 
irrelvant to our study in this work.}, following Ref.~\cite{maximal}. 
The Ga data requires $\nu_e$ survival
probability of $\sim$ 5/9, which implies $L\over E$ for the solar neutrino
is smaller than $1\over (\Delta m^2_{ij})_{\rm min}$.  The atmospheric
neutrino data and km range laboratory experiments require another
$\Delta m^2_{ij}$. Hence, two mass difference scales have been all used, and 
there is none left for a new scale suggested by the LSND data around
$\Delta m_{LSND}^{2} \sim O(1)~{\rm eV}^2$ with a mixing angle 
$\sim ({\rm a ~few }) \times 10^{-3}$.
The only possibility
to explain both the mass shifts at LSND point and at atmospheric data points
in the maximal mixing scenario is that there are two thresholds corresponding 
to a larger $\Delta m^2$
at $\sim O(1)$ eV$^2$ and a smaller $\Delta m^2$ at around $10^{-2}$ eV$^2$.
In this case, the $\nu_e$ survival probability for the solar neutrino 
problem is $1/3$, and is too small to accommodate the Ga data.  
Therefore, although the qualitative features of the maximal mixing scenario 
is encouraging, it is not viable if the LSND data is confirmed in the 
future.  Another way to see this is as follows : the maximal mixing scenario
predicts the transition probability for $\nu_{\mu} \rightarrow \nu_{e}$ 
to be 4/9 in the range of the 
LSND experiment, which clearly contradicts  the reported transition
probability, (a few) $ \times 10^{-3}$.   
Furthermore, the best $\chi^2$ fit to the atmospheric and the solar 
neutrino data indicates that the masses of three light neutrinos are 
$m_{3} \simeq (85 \pm 10)$ meV, and $m_{1,2} < 3 \mu$eV \cite{maximal}, 
which are too 
light to be  cosmologically interesting as a hot dark matter component of
the missing mass of the universe. 

Therefore, we make an ansatz for the neutrino mass matrix 
which is a simple modification of the maximal mixing one, (1), and 
study its consequences in this work.
We assume that neutrinos are Dirac particles so that the 
lepton number is to be conserved in our model. 
Then, each left-handed neutrino ($\nu_{L}^i)$ is accompanied by 
the right-handed partner ($\nu_{R}^i$) which is sterile under electroweak 
interactions.
 
Note that any $3 \times 3$ matrix $M_{ij}$ can be decomposed 
as $M = X - i Y$ with both $X$ and $Y$ hermitian.
Also any hermitian matrix 
$X$ can be written as $ X = S + i A$, where $S$ ($A$) is a real 
(anti)symmetric matrix. Finally, the symmetric matrix $S$ can be decomposed
as the trace part proportional to $\delta_{ij}$ and the traceless symmetric
matrix.
One can combine the trace part of the symmetric mass matrix $S$ and the real
antisymmetric part $A$ in order  to get the neutrino mass
matrix, 
\begin{eqnarray}
M = \mu~\left( \begin{array}{ccc}
             1 & ic & i d    \\
             -ic & 1 & ib \\
             -i d & -ib & 1
           \end{array} \right),
\label{eq:ourm}
\end{eqnarray}
where $\mu$ is the mass scale,  $b,c$ and $d$ are all real. 
We have chosen a basis in which the charged lepton mass 
matrix is diagonal.  Note that the diagonal terms are still universal, and 
only the off-diagonal elements are modified from the maximal mixing one,
(1).   Note that this ansatz becomes the maximal mixing one if $b=c=-d$ 
({\it i.e.}, if there is a permutation symmetry among three generations).
This form for the mass matrix is sufficiently simple but rich enough to
give nontrivial analytic formulae for the survival and transition 
probabilities for three neutrinos. 
If there is a solution when two of the off-diagonal elements are the same
(say, $b = d$ for example),
then the permutation symmetry among three generations ($S_3$) 
in  the original
maximal mixing case (1) breaks down to $S_2$. This would imply that
our mass matrix ansatz depends on three real parameters and, thus it 
is {\it one of the simplest modifications to the maximal  mixing ansatz},
which can accommodate LSND, atmospheric and solar neutrino data. 
In fact, this is the case (with $b^{2} = d^{2}$) as discussed in the 
following.  
It breaks the original $S_3$ possessed by (1) into $S_2$ in a particular
way. There may be several other ways to break this permutation symmetry,
which will be considered elsewhere.
There may be some underlying dynamical reasons for the above 
form of the mass matrix,
but we take it as  a simple phenomenological ansatz for the moment.

Three eigenvalues of the mass matrix (3) are 
\begin{equation}
m_{1} = \mu, ~~~~~~~~   m_{2,3} =  \mu ( 1 \pm N),
\label{eq:mass}
\end{equation}
where $N = ( b^{2} + c^{2} + d^{2} )^{1/2}$.  
The corresponding eigenvectors 
form the mixing matrix $U$ which relates the weak eigenstate $\nu_{\alpha}$ 
to the mass eigenstates $\nu_{i}$ as $\nu_{\alpha} = \Sigma_{i} U_{\alpha i} 
\nu_{i}$. 
The indices $\alpha = e, \nu, \tau$ label the flavor eigenstates,
and $i = 1,2,3$ label the mass eigenstates of three neutrinos.

Then one can easily verify that
\begin{eqnarray}
P( \nu_{e} \rightarrow \nu_{e} ) & = & 1 - {1\over N^4}~\left[ \left(
\Delta_{21} + \Delta_{31} \right) b^{2} (N^{2} - b^{2}) + {1\over 2}~
\Delta_{32}~(N^{2} - b^{2})^{2} \right],  
\label{eq:pee}
\\
P( \nu_{\mu} \rightarrow \nu_{e} ) & = & {1 \over N^4}~\left[ ( \Delta_{21}
+ \Delta_{31} ) b^{2} d^{2} + {1 \over 2}~\Delta_{32} ( c^{2} N^{2} - 
b^{2} d^{2} )    \right], 
\label{eq:pem}
\\
P(\nu_{\mu} \rightarrow \nu_{\mu}) & = & 1 - {1 \over N^4}~ \left[
( \Delta_{21} + \Delta_{31} ) d^{2} ( N^{2} - d^{2} )       \right.
\label{eq:pmm}
\\
& & \left. + \Delta_{32} \left\{ \left( {{c^{2} N^{2} - b^{2} d^{2}} 
 \over
2}\right) + {{(c^{2}N^{2} + b^{2} d^{2} ) ( b^{2} c^{2} + N^{2} 
d^{2} )}
 \over
2 ( c^{2} + d^{2})^2} - c^{2} d^{2} \right\} \right],
\nonumber  
\\
P(\nu_{\mu} \rightarrow \nu_{\tau}) & = & { 1 \over N^4}~\left[ 
( \Delta_{21} + \Delta_{31} ) c^{2} d^{2} + \Delta_{32} \left( 
{{(c^{2} N^{2} + b^{2} d^{2} )(b^{2} c^{2} + d^{2} N^{2} )} \over 
{2 (c^{2} + d^{2})^2}} - c^{2} d^{2} \right) \right],
\label{eq:pmt}
\end{eqnarray}
where 
\begin{equation}
\Delta_{ij} = 2~\sin^{2} \left( 1.27 ~L ~\Delta m^2_{ij} \over E \right),
\end{equation}
with $\Delta m^2_{ij} \equiv m_{i}^{2} - m_{j}^2$  in eV$^2$
and $L/E$ in km/GeV. 
Since $\sum\Delta m^2_{ij}=0$, 
there exist only two independent mass difference parameters.

Note that the heights of the plateaus for the $\nu_e$ survival probability are 
functions of $b^{2} / N^2$ only, even in the presence of nonvanishing $c$ 
and $d$.
Since $P(\nu_{\mu} \rightarrow \nu_{e})$ depends on two mass 
differences, one can identify the mass difference $\Delta m_{LSND}^{2} 
\sim O(1) ~{\rm eV}^{2}$ either as $\Delta m_{31}^2$ 
or as $\Delta m_{32}^2$. 
The other mass difference is taken to be $0.72 \times 10^{-2}~{\rm eV}^2$
in order to solve the atmospheric neutrino problem.  Thus, there is a 
reasonable hierarchy between two mass differences.
In the following, we discuss two possibilities separately.

(I) : $\Delta m_{31}^2 = \Delta m_{LSND}^{2} \sim O(1)~{\rm eV}^{2}$ 
and $\Delta m_{32}^2 = 0.72 \times 10^{-2}~{\rm eV}^2$ : 

In this case, the heights of the intermediate plateaus for the 
$\nu_{e}$ and $\nu_{\mu}$ survival probabilities are given by
\begin{eqnarray}
P(\nu_{e} \rightarrow \nu_{e}) & = & 1 - 2 {b^{2} \over N^2}~\left(1 - 
{b^{2} \over N^2} \right),
\\
P( \nu_{\mu} \rightarrow \nu_{\mu} ) & = & 1 - 2 {d^{2} \over N^2}~\left(
1 - {d^{2} \over N^2} \right),
\end{eqnarray}
and the transition probabilities can be approximated as
\begin{eqnarray}
P( \nu_{\mu} \rightarrow \nu_{e}) & = & 4 { b^{2} d^{2} \over N^4} 
\sin^{2} \left( {1.27 L \Delta m_{31}^{2} \over E } \right),
\\
P( \nu_{\mu} \rightarrow \nu_{\tau}) & = & 4 {c^{2} d^{2} \over N^4}
\sin^{2} \left( {1.27 L \Delta m_{31}^{2} \over E } \right).
\end{eqnarray}
The transition probability is often described in terms of two parameters,
$\Delta m^2$ and $\theta_{\alpha \beta}$ for which
\begin{equation}
P ( \nu_{\alpha} \rightarrow \nu_{\beta} ) = \sin^{2} 2 \theta_{\alpha 
\beta} ~\sin^{2} \left( {1.27 L \Delta m_{31}^{2} \over E } \right).
\end{equation}
Therefore, we can make the following identifications :
\begin{eqnarray}
\sin^{2} 2 \theta_{e \mu} & = & {4 b^{2} d^{2} \over N^4},
\\
\sin^{2} 2 \theta_{\mu \tau} & = & {4 c^{2} d^{2} \over N^4},
\end{eqnarray}
in our model (for $\Delta m_{31}^{2} >> \Delta m_{32}^2$).  

Experimental  results for the neutrino oscillations are shown in the
($\sin^{2} 2 \theta, \Delta m^2$) plane. 
In the plot presented by the LSND group, there are small regions in 
this plane which 
indicates a possible transition of $\bar{\nu}_{\mu} \rightarrow \bar{\nu}_e$.
This region is not ruled out by other laboratory searches such as 
BNL E776 \cite{bnl}, KARMEN \cite{karmen}, BUGEY \cite{bugey} and others
\cite{gosgen}-\cite{krasnoyarsk}.  For each possible $\Delta m_{LSND}^2$, 
we show the 
possible value(s) of $\sin^{2} 2 \theta_{e\mu}$ in Table~1.  For the same
$\Delta m_{LSND}^2$, there is an upper bound on $\sin^{2} \theta_{\mu 
\tau}$ from 
FNAL E531 \cite{fnale531}, CHARM II \cite{charmii} and CDHSW \cite{cdhsw}, 
and we also list these numbers in Table~1.

For each $\Delta m_{31}^2 = \Delta m_{LSND}^2$ given in Table~1, 
one can solve Eq.~(4) to get
the neutrino masses.  For example, $\Delta m_{31}^2 = 6~{\rm eV}^2$ leads to
\begin{eqnarray}
m_{1} & = & 7.35 \times 10^{-4}~~{\rm eV},
\nonumber   \\
m_{2} & \approx& -  m_{3} \approx 2.45~~{\rm eV},
\label{eq:numass}
\end{eqnarray}
with $\Sigma_{i} |m_{\nu_{i}} | = 4.9$ eV.  
(The negative $m_3$ can be remedied by a chiral transformation of
$\nu_3$ field.)  For other values of $\Delta m_{31}^2$,  we show the resulting
neutrino masses from our mass matrix ansatz (3) in the fourth column of
Table~1.

These light neutrinos can contribute to the missing mass of 
the universe (the hot dark matter) in amount of \cite{kolb} 
\begin{equation}
\Omega h^{2} = 7.83 \times 10^{-2}~{g_{eff} \over g_{*s}(T_{D})}~\left(
{m_{\nu} \over eV } \right),
\label{eq:omega}
\end{equation}
where $g_{*s}(T_{D}) = 10.75$ and $g_{eff} = (3 g)/4 = 3/2$  are the 
effective degrees of freedom contributing to the entropy density 
$s$ and to
the ratio $Y = n/s$, $n$ being the number density,
respectively. 
The parameter $h$ is related to the Hubble 
constant $H_{0}$ as $H_{0} = 100 h ~{\rm km~ sec}^{-1}~{\rm Mpc}^{-1}$ 
\cite{kolb}.  
So, for the solution (17), three light 
neutrinos can constitute 8.3 \% ~(21.4 \%) of the  missing mass of the 
universe for $h = 0.8~ (0.5)$, which is again cosmologically interesting.  
(Here, we have assumed that three sterile right-handed neutrinos 
decouple much earlier than the left-handed 
neutrinos, and that they don't affect the results 
of the standard cosmology.)  The results for other values of $\Delta m_{31}^2$
are listed in the last column of Table~1.  

When we determine $b^{2}$ and $d^2$, it is important to satisfy all the 
constraints shown in Table~1.  
One might try to perform the $\chi^2$ fit to the available data on the
neutrino oscillations.  Instead, we choose to scan $d^2/N^2$ for each 
$\Delta m^2$ in the first column of 
Table~1.  For each $d^2/N^2$, the parameter $b^2$ is determined by the 
mixing angle given by
the LSND experiments, and $c^{2}/N^{2} = (1 - b^{2}/N^{2} - d^{2}/N^{2})$.  
Then, we require that the resulting $\sin^{2} 2 \theta_{\mu \tau}$ satisfy
the upper limit given in the third column of Table~1.  We also calculate
the survival probabilities for $\nu_e$ and $\nu_{\mu}$ at the intermediate 
level (the laboratory and the km range scale) and require them to be larger 
than 0.95 in order to satisfy the null results in various types of 
disappearance experiments for the $\nu_{\mu}$ and $\bar{\nu}_e$ beams.  
For $\Delta m^{2} = 6~{\rm eV}^2$ or larger (the first and the second rows),
the resulting $d^{2}/N^{2} \approx 0.995-1.00$, which corresponds to 
almost no disappearance of $\nu_{\mu}$ for all ranges of $L$ and $E$.  
Thus, we reject $d^{2}/N^2$ around 1.  
For $\Delta m_{31}^{2} \le 2~{\rm eV}^2$, the allowed ranges for $d^{2}/N^2$ 
are typically around 0.010--0.020.  
The corresponding $b^{2}/N^2$'s are also in the same range as $d^{2}/N^2$.
Thus, as discussed in the following,  we can accommodate
the laboratory scale and the large scale neutrino experiments, 
by choosing small (but nonvanishing) $b^2$ and $d^2$. 

In particular, there is a small region in which $b^{2} = d^{2}$  gives
acceptable fits to all available data on neutrino oscillation 
experiments except for the $^{37}$Cl data.  This is quite interesting, since
it corresponds to residual permutation symmetry ($S_{2}$)
between $e$ and $\mu$ in (3) with three real parameters. 
In other words,  we have  
$ | M_{e \mu} | = | M_{\mu \tau} |$.  Our matrix with $b^{2} = d^2$
breaks the original
symmetry of the maximal mixing one ($S_{3}$) into $S_{2}$, and thus may be 
regarded as one of the simplest modifications to the maximal mixing one, (1).
In the following, we demonstrate that the ansatz
(3) can  reasonably fit all the data with $ b^{2}/N^{2} = 
d^{2} /N^{2} = 0.015$ for $\Delta m_{31}^{2} = 2~{\rm eV}^2$,
except for the HOMESTAKE data, with a reasonable accuracy at present. 

In Figure 1, we show the resulting survival probability for 
$\nu_e$ in the solid curve 
\footnote{ In order to average the 
oscillation probabilities, we adopt the
prescription by Harrison {\it et al.} \cite{maximal}, which amounts to
replacing $\cos (x/2)$ by $\sin x / x$.} along with various 
types of neutrino oscillation data, the $\nu_{e}$ disappearance 
experiments at reactors  \cite{karmen}-\cite{krasnoyarsk} and 
the solar neutrino experiments \cite{kamioka2}-\cite{gallex}.  
Using three neutrino mixing with four parameters, we get two
step survival probability for $\nu_{e} \rightarrow \nu_e$.     
The plateau for the large $L$ is about 0.49, a little
bit lower than $5/9$ used before in the maximal mixing case. 
Thus, the solar neutrino deficit is solved in terms of vacuum oscillations.
The intermediate plateau of $\sim 0.97$ for $\nu_e$ is the prediction of our
specific form of the mass matrix, and is consistent with all the existing 
data.  The KRASNOYARSK data has a relatively large error bar, and our curve
is within two $\sigma$ of the data.   

In Figure 2, the survival probability for $\nu_{\mu}$ using our ansatz is 
shown in the solid curve, along with the $\nu_{\mu}$ disappearance 
experiment data \cite{cdhs} \cite{charmps}. The survival  
probability for $\nu_{\mu}$ at large $L$ is $0.48$, and the intermediate 
plateau has a height of $0.97$.  
The curve agree with the available data quite well.   

For the atmospheric neutrino data, we show the data point for the so-called
$R$ defined by
\begin{equation}
R \equiv {{( N_{\mu} / N_{e} )_{Data}} \over {( N_{\mu} / N_{e} )_{MC} }},
\end{equation}
along with our prediction 
\begin{equation}
R = {{P_{\mu \mu} + P_{e \mu} / r} \over  {P_{ee} + r P_{\mu e} }}
\end{equation}
in Table~2, where $r$ is the incident $(\mu / e)$ ratio.   From 
Table~2, we observe that most of the predicted $R$ values are consistent
with  all the atmospheric neutrino data from KAMIOKANDE, IMB, and others
\cite{kamioka}-\cite{soudan}, considering some  data points have large
errors.

For these numbers with $\Delta m_{31}^{2} = 2~{\rm eV}^2$, 
we predict $\sin^{2} 2 \theta_{\mu \tau} \approx  6 \times 
10^{-2}$, which is just below the current upper limit, $8 \times 10^{-2}$.
This range of $\sin^{2} 2 \theta_{\mu \tau}$ may be probed at 
CHORUS, NOMAD, FNAL P803, CERN/ICARUS and FNAL/SOUDAN2 \cite{gelmini}.
It would be interesting to test our predictions for the $\nu_{\mu} 
\rightarrow \nu_{\tau}$ oscillation in the future.  
Similar results can be drawn for other values of $\Delta m_{31}^2$.
Thus, our mass matrix ansatz (3) not only describes all the available data
on neutrino oscillations, but also predicts the mixing angle for $\nu_{\mu}
\rightarrow \nu_{\tau}$ in an interesting range which lies within 
sensitivity of the near-future experiments. 

Let us briefly discuss the second case : (II) $\Delta m_{31}^{2} = 0.72 
\times 
10^{-2}~{\rm eV}^2$, and $\Delta m_{32}^{2} = \Delta m_{LSND}^2$.
In this case, it is easy to verify that there is no solution for 
$b^{2}/N^{2}$ and $d^{2}/N^2$ which satisfy the constraints from the 
laboratory scale experiments from BUGEY, BNL E766, and those in the second
and the third columns of Table~1. 
So, our mass matrix ansatz prefer the solution (I) for which $m_{1}$ is 
smallest around $10^{-3}-10^{-4}$ eV, and the other two are nearly degenerate
with $m_{2} \approx m_{3} \approx O(1)$ eV.   

In summary, the neutrino mass matrix ansatz (3)  with four 
real parameters is one of the simple modifications to the maximal mixing 
ansatz (that
fails to fit the new LSND data)   which can 
fit various types of neutrino experiments except for the $^{37}$Cl 
solar neutrino data \cite{homestake}. 
In particular, there are solutions with $b^{2} = d^{2}$ with residual
permutation symmetry among two generations ($\nu_{\mu} \leftrightarrow 
\nu_{e}$).  In this sense, our mass matrix ansatz could  be regarded as
one of the simplest modifications to the maximal mixing ansatz.  
The resulting light neutrinos have masses (17), and thus they   
can constitute about 2.4--4.8 \%~ (6.2--12.4 \%) of the missing mass of 
the universe for $h = 0.8 (0.5)$, and thus cosmologically interesting
unlike the maximal mixing scenario.
In our model, the transition probability for $\nu_{\mu} \rightarrow 
\nu_{\tau}$ is close to the current upper limit, depending on $\Delta 
m_{LSND}^2$ as shown in the third column of Table~1.
Since it lies within the reach of 
various future experiments such as CHORUS, NOMAD, {\it etc.}, the observation
of $\nu_{\mu} - \nu_{\tau}$ oscillation would constitute a definite test 
of our mass matrix ansatz along with the confirmation of the LSND data.  

\acknowledgements

We thank Bob Lanou for helpful discussions on the LSND and other 
neutrino experiments, and Dr. Hang Bae Kim for his assistance to draw 
the figures. One of us (JEK) thanks the Brown High Energy Theory Group for 
the hospitality extended  to him during the visit.
This work is supported in part by the Korea Science and Engineering 
Foundation through Center for Theoretical Physics at Seoul National
University (JEK, PK), SNU-Brown Exchange Program (KK, JEK),
Korea-Japan Exchange Program (JEK), the
Ministry of Education through the Basic Science Research Institute,
Contract No. BSRI-94-2418 (JEK) and through Contract No. BSRI-94-2425 (PK),
SNU Daewoo Research Fund (JEK), and also the US DOE Contract DE-FG-02-
91ER40688 - Task A (KK).

\vspace{0.5in}

{\it Note Added in Proof}
\vspace{.2in}

In Eqs.~(5)-(8), we did not show terms involving $s_{ij} \equiv 
\sin \left( {{1.27 L  \Delta m_{ij}^2 } \over E } \right)$, since these
terms vanish under taking averages, or they are irrelvant to our study. 
See Ref.~[5] for more details.  


%
%
\begin{figure}
\caption{The survival 
probabilities $P( \nu_{e} \rightarrow \nu_{e} )$ using our 
ansatz (3) for the case (I),
along with the reactor experiment data from KARMEN, ILL/GOSGEN, BUGEY, 
KRASNOYARSK, and the solar neutrino data from KAMIOKA, HOMESTAKE, SAGE 
and GALLEX.  }
\label{figone}
\end{figure}

\begin{figure}
\caption{The survival 
probabilities of $\nu_{\mu}$ using our ansatz (3)
for the case (I),
along with the accelerator experiments from CDHS-SPS and CHARM-PS.}
\label{figtwo}
\end{figure}




%
%
\begin{table}
\caption{ The allowed regions for $\Delta m^2$ and $\sin^{2} 2 \theta_{e\mu}$ 
consistent with the LSND as well as BNL E766, and the corresponding upper 
limit for $ \sin^{2} 2\theta_{\mu \tau}$ from FNAL E531 and CDHSW. 
The fourth column is the predicted neutrino masses by our mass matrix ansatz
(3). The last column shows contributions of three light neutrinos to the
missing mass of the universe for $h = 0.8 ~(0.5)$.  See the text for details. 
}
\label{table0}
\begin{tabular}{c|cccc}
$\Delta m^2~({\rm eV}^2) $ & $\sin^{2} 2 \theta_{e\mu}$ & $ \sin^{2} 2 
\theta_{\mu \tau}$  &   $ (m_{1}, m_{2}, m_{3})~{\rm in~ eV} $ 
& $\Omega$ (\%)  \\    \tableline
 20    &  $ \sim 3 \times 10^{-3}$   &  $ < 4 \times 10^{-3}$
& $( 4.02 \times 10^{-4}, 4.47, -4.47)$ & $15.1 \%~(38.9 \%)$ \\
6       &  $ 2 \times 10^{-3}$ & $ <  2 \times 10^{-2}$ & $(7.35 \times 10^{-4},
2.45, -2.45)$ & $8.3 \% ~(21.4 \%)$ 
\\
2       &  $ (1\sim 2) \times 10^{-3}$   & $ < 8 \times 10^{-2}$
& $(1.27 \times 10^{-3}, 1.43, -1.43)$ & $4.8 \% ~(12.4 \%)$ \\
1       &  $ (2 \sim 6) \times 10^{-3}$   & $ < 0.1$
& $ (1.80 \times 10^{-3}, 1.0, -1.0)$  & $ 3.4 \% ~(8.8 \%)$  \\
0.5     &  $ (0.7 \sim 2) \times 10^{-3}$   & $ < 0.2$
& $(2.55 \times 10^{-3}, 0.71, -0.71)$ & $2.4 \%~(6.2 \%)$    
\end{tabular}
\end{table}

\begin{table}
\caption{The atmospheric neutrino data $R$ for various $L/E$ along with
our predictions for $\Delta m_{31}^{2} = 2~{\rm eV}^2$, $\Delta m_{32}^{2} = 
0.72 \times 10^{-2}~{\rm eV}^2$ and $b^{2}/N^{2} = d^{2}/N^{2} = 
0.015$.  We show the $r = (\mu/e)_{\rm incident}$ 
values for each data point also.}
\label{table1}
\begin{tabular}{ccccc}
Experiments & $r$ & $L/E$ (km/GeV) & Measured & Prediction 
\\   \tableline
KAMIOKA \cite{kamioka} & $4.5/1$ & 5  & $1.27^{+0.61}_{-0.38}$ & 
0.99
\\
(Multi-GeV)            & $3.2/1$ & 10 & $0.63^{+0.21}_{-0.16}$ & 
0.97
\\
		       & $2.2/1$ & 100& $0.51^{+0.15}_{-0.12}$ & 
0.41
\\
                       & $3.2/1$ & 1000 & $0.46^{+0.18}_{-0.12}$ & 
0.31
\\
                       & $4.5/1$ & 2000 & $0.28^{+0.10}_{-0.07}$ & 
0.22
\\
KAMIOKA \cite{kamioka} & $2.1/1$ & 80   & $0.59 \pm 0.10$ & 
0.50
\\
(Sub-GeV)              & $2.1/1$ & 12800 & $0.62 \pm 0.10$  &
0.48
\\
IMB \cite{imb}         & $2.1/1$ & 1000 &  $0.54 \pm 0.13$ & 
0.47
\\
FREJUS \cite{frejus}   & $2.1/1$ & 500  &  $0.87 \pm 0.18$ &   
0.47
\\
NUSEX \cite{nusex}     & $2.1/1$ & 500  &  $0.99 \pm 0.32$ & 
0.47
\\
SOUDAN \cite{soudan}   & $2.1/1$ & 1000 & $0.69 \pm 0.21$  & 
0.47
\end{tabular}
\end{table}

\end{document}